\documentclass[aps,pre,twocolumn,superscriptaddress]{revtex4}
\usepackage{graphics, graphicx, epsfig}

\begin{document}

\title{Quantifying the Bicoid morphogen gradient in living fly embryos}
\author{Julien Dubuis}
\affiliation{Joseph Henry Laboratories of Physics, Princeton University, New Jersey 08544, USA}
\author{Alexander H Morrison}
\affiliation{Joseph Henry Laboratories of Physics, Princeton University, New Jersey 08544, USA}
\author{Martin Scheeler}
\affiliation{Joseph Henry Laboratories of Physics, Princeton University, New Jersey 08544, USA}
\author{Thomas Gregor}
\affiliation{Joseph Henry Laboratories of Physics, Princeton University, New Jersey 08544, USA}
\affiliation{Lewis Sigler Institute for Integrative Genomics, Princeton University, New Jersey 08544, USA}

\begin{abstract}

In multicellular organisms, patterns of gene expression are established in response to gradients of signaling molecules. During fly development in early Drosophila embryos, the Bicoid (Bcd) morphogen gradient is established within the Þrst hour after fertilization. Bcd acts as a transcription factor, initiating the expression of a cascade of genes that determine the segmentation pattern of the embryo, which serves as a blueprint for the future adult organism. A robust understanding of the mechanisms that govern this segmentation cascade is still lacking, and a new generation of quantitative measurements of the spatio-tempral concentration dynamics of the individual players of this cascade are necessary for further progress. Here we describe a series of methods that are meant to represent a start of such a quantification using Bcd as an example. We describe the generation of a transgenic fly line expressing a Bcd-eGFP fusion protein, and we use this line to carefully analyze the Bcd concentration dynamics and to measure absolute Bcd expression levels in living fly embryos using two-photon microscopy. These experiments have proven to be a fruitful tool generating new insights into the mechanisms that lead to the establishment and the readout of the Bcd gradient. Generalization of these methods to other genes in the Drosophila segmentation cascade is straightforward and should further our understanding of the early patterning processes and the architecture of the underlying genetic network structure.

\end{abstract}

\maketitle

\section{Introduction}

Patterning of multicellular organisms results from the interpretation of morphogen gradients by small genetic regulatory networks. The inputs and outputs of these networks are protein molecules that are synthesized by the cell and that act as Òtranscription factorsÓ which bind to the DNA to control downstream network elements. Quantitatively mapping the relationships between inputs and outputs as well as characterizing the noise of these regulatory elements are essential for our global understanding of the patterning network. Over the past decade, a physical picture of the noise in genetic control \cite{Blake03,Elowitz02,Ozbudak02,Pedraza05,Raser04} and of the global network structure that patterns the embryo \cite{Fujioka99,Jaeger04a,Peter09,Reinitz95} has been fairly well established. Therefore, we can use this knowledge to ask questions about the overall function and design of such networks as well as about their capacity to transmit positional information, the knowledge individual cells acquire about their spatial location within the organism. 
Our current understanding of such networks is mainly derived from genetics and static images of fixed tissue \cite{Jaeger04b}. To fully describe the spatiotemporal regulatory interactions that determine patterning, however, a complete dynamic view is needed. Development is an intrinsically dynamic process where spatial and temporal components are intimately tied together. Characterizing the dynamics of development is important both for gaining concrete visual insights into complex developmental processes and for testing the plausibility of simple mechanisms implied by proposed models for gradient formation \cite{Crick70, Kicheva07, Bergman07, Coppey07,DeLotto07,Hecht09} and gene regulation \cite{Dassow00,Manu09,Bialek05,Tostevin09}.

Furthermore, for a fully quantitative understanding of the genetic regulation that determines the early patterning processes, we need to be able to make high precision measurements of the relevant protein concentrations in living embryos. Such measurements are nontrivial because they require high image resolution, high sensitivity, and small errorbars, which are achieved best through larger light source intensities and slow acquisition modes. However, high energies result in photobleaching of the specimen and slow acquisition times are incompatible with the inevitable concentration changes intrinsic to development. Overexposure of the embryo with light energy might interfere with the measured quantity and with the natural course of development.
Finally, it is important to determine the correct correlation between the number of photons collected and the protein concentration being measured. In order to give a quantitative confidence to the measured protein concentrations, careful errorbar estimation is necessary.

\section{Experimental apparatus}

\begin{figure} 
   \includegraphics[width=3.2in]{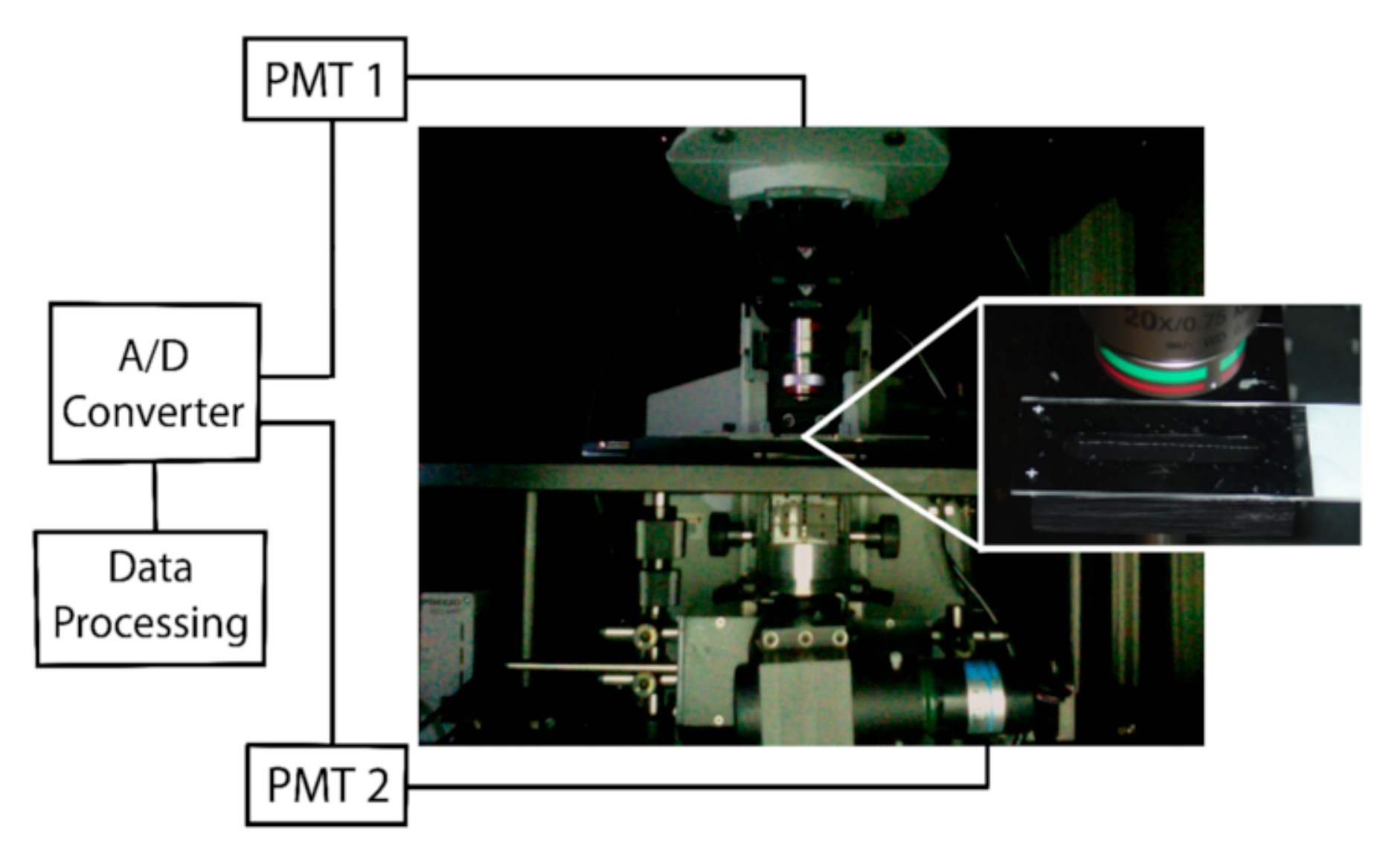}
   \centering
   \caption{Two-photon microscope setup. An image of the objective, the stage, and a high NA oil-immersion condenser with trans- and epi-detection systems highlighted. Insert shows a typical slide of embryos for mounting.}
   \label{Figure1}
\end{figure}

A custom-built two-photon excitation laser scanning microscope \cite{Denk90} is used for all in vivo imaging of Drosophila embryos. The microscope is comprised of a combination of commercial and custom parts, adapted to increase light collection through simultaneous detection of both epi- and transflorescence \cite{Mainen99,Svoboda97}. Figure~\ref{Figure1} depicts the objective, stage, and condenser of the experimental apparatus. Samples are excited by light from a mode-locked pulsed Ti:Sapphire laser (Mira 900, Coherent, $\sim100$-fs pulses at 80 MHz), whose wavelength is tuned to $\sim 920$ nm by a custom set of Mid-band filters. The laser power can be varied with an opto-electric light modulator or Pockels cell (Conoptics, model 350-80LA). Coupled scanning mirrors are used to keep the beam stationary with respect to the stage, which is capable of translation in the X, Y, and Z directions through the use of a modified Sutter MP285 micromanipulator. Detection efficiency is increased through the use of both a condenser and a $25\times$, 0.8 Imm Korr DIC objective (Zeiss), whose signals are then amplified by two separate PMTs (Hamamatsu R3896 and C6270). The advantage of utilizing this additional trans-fluorescence lies in the larger NA of the oil-immersion condenser (NA=1.4) which provides an increased collection efficiency resulting in an increased signal-to-noise ratio by a factor of $ \sim2$. This configuration also prevents loss of collection efficiencies due to scattering, as the signal loss in the epi-channel sustained with increased tissue depth is compensated by the corresponding increase in the trans-channel signal, leaving the sum of epi- and transfluorescent signals approximately constant with variations in tissue depths \cite{Mainen99}. The microscope is controlled by customized ScanImage software \cite{Pologruto03}, which is also used in managing the specifications of each acquisition.

\section{Fly strain generation}

The Bicoid (Bcd) morphogen gradient in the fruit fly Drosophila melanogaster is an ideal system to start deciphering in vivo transcription factor dynamics. It is the primary input into the network that determines anterior patterning of the early embryo; a qualitative static picture of its form and function is very well established \cite{Nusslein80,Frohnhofer86,Nusslein87,Driever88a,Driever88b,Driever89,Struhl89}; and the major part of its activity only happens more than an hour after its expression begins, leaving sufficient time for maturation processes of the relevant proteins to complete. 
To visualize the spatiotemporal dynamics of Bcd, transgenic Drosophila embryos have been made in which endogenous Bcd was replaced with a green ßuorescent fusion protein (called Bcd-GFP hereafter) \cite{Gregor07b}. In order to ensure the biological relevance of protein level measurements made with this fly strain, its Bcd expression levels need to resemble as closely as possible the endogenous wild type levels. This is particularly important in high precision measurements of low protein levels and of their fluctuations. To generate the fly strain we used a plasmid carrying a transcript that encodes for a recombinant Bcd protein fused to eGFP (enhanced Green Fluorescent Protein) \cite{Tsien98, Mavrakis10} at its N terminus. The fusion construct \cite{Hazelrigg98} had a size of 6.5 kb and contained endogenous bcd 5Õ and 3Õ UTRs, which are known to mediate anterior localization and translation of bcd mRNA. This construct completely rescues embryos from bcd mutant mothers: qualitatively, no developmental defects are detected throughout the entire life cycle, and quantitatively, measured cues that directly follow from the embryoÕs biological and physical properties are identical to wild type: the position of the cephalic furrow and the gradientÕs length constant \cite{Gregor07b}. The latter measures ensure that both the protein concentration levels and the protein dynamics mimic their natural counterparts in wild type embryos, justifying the relevance of their subsequent quantification. 

\section{Linearity of antibody stainings}

\begin{figure} 
   \includegraphics[width=3.2in]{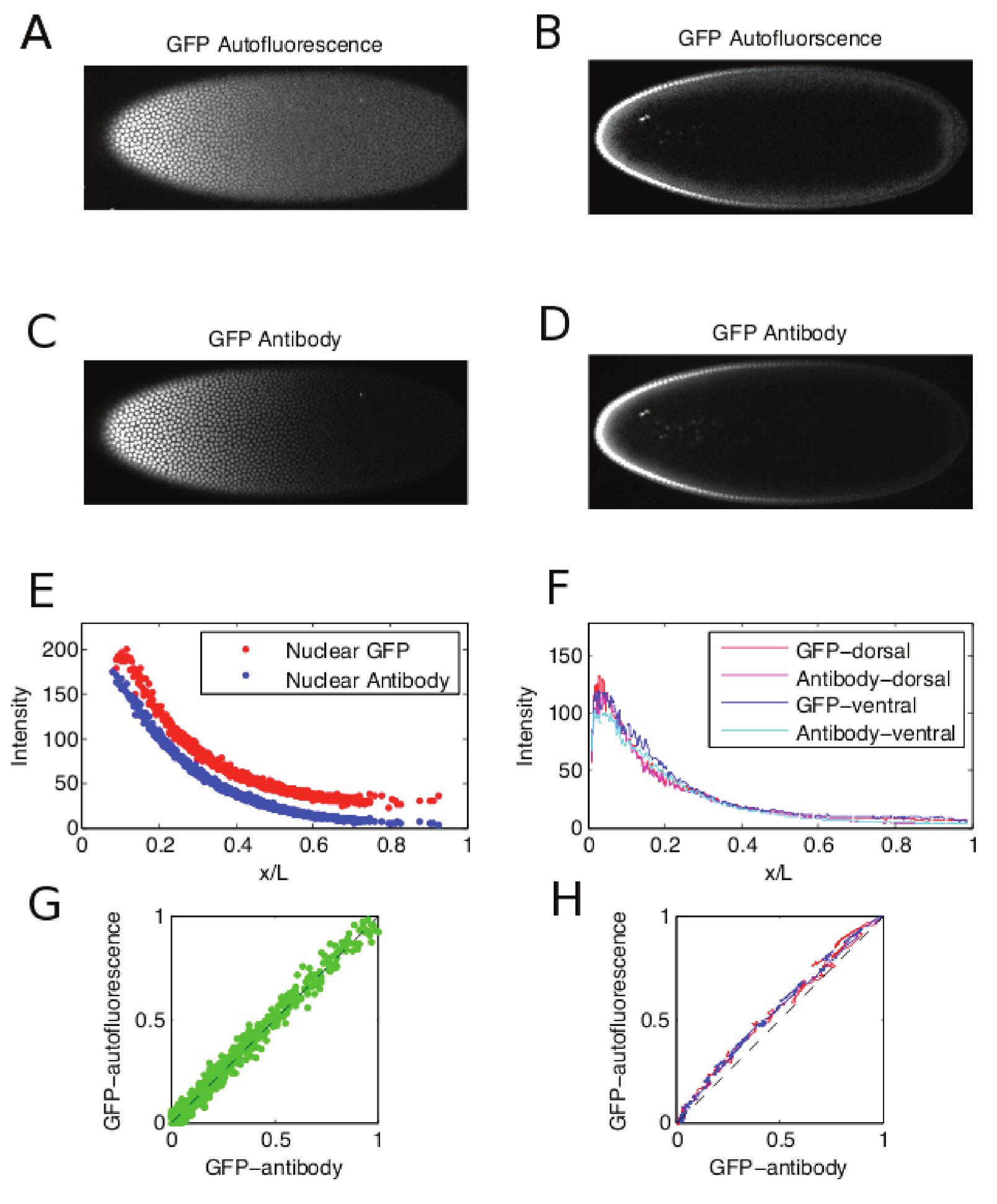}
   \centering
   \caption{Linearity of antibody stainings. \textbf{A-D} A single embryo was formaldehyde fixed during nuclear cycle 14 (for 45 minutes in 6.7\% paraformaldehyde, $1\times$ PBS) and stained with rabbit anti-GFP primary antibody (Chemicon) following previously published protocols \cite{Wieschaus86}. The secondary antibody was conjugated with an infra-red Alexa-647 (Molecular Probes), maximally reducing spectral overlap with the green GFP-autofluorescence. The embryo was imaged both at the surface (A and C) and the midsagittal plane (B and D) via confocal microscopy (Leica SP5, $20\times$ oil immersion objective plan apochromat (Leica, NA=0.7)). GFP-autofluorescence (A and B) and anti-GFP staining (C and D) were recorded in consecutive runs. \textbf{E} Extracted raw fluorescence intensity profiles from A and C projected on the embryoÕs AP axis. Each point corresponds to a single nucleus, for details see \cite{Gregor07b}. \textbf{F} Raw fluorescence intensity profiles from B and D projected on the embryoÕs AP axis, extracted by sliding a circular averaging area along the edge of the embryo, see \cite{Houchmandzadeh02} for details. \textbf{G} Scatter plot of GFP-autofluorescence intensities vs. fluorescence antibody staining intensities extracted from A and C for all nuclei. Each point corresponds to a single nucleus, curves normalized by nuclei of maximal and minimal intensities. \textbf{H} Scatter plot of fluorescence intensities extracted from B and D. The blue line corresponds to dorsal profile while the red line corresponds to ventral profile.}
   \label{Figure2}
\end{figure}

Previously, gene expression levels in Drosophila embryos had been quantified using fluorescent antibody stainings \cite{Houchmandzadeh02,Jaeger04b,Kosman98}. Such quantification relies on the assumption that the fluorescence intensity levels extracted from stained embryos and the actual protein concentrations detected by the antibodies are linearly related to the embryoÕs natural protein concentration levels. The Bcd-GFP fusion construct allows for a direct test of this linearity of antibody stainings as a method of quantifying relative protein concentrations. This can be achieved by staining Bcd-GFP embryos with an antibody against GFP (or Bcd) and simultaneously measuring the autofluorescence of GFP and the intensity of the antibody staining.  The principal difficulty here is to avoid damaging the Bcd-GFP protein during the staining protocol. So, in order to avoid the severe attenuation of the GFP auto-fluorescence by the usual methanol treatment during the fixation process, embryos were fixed in paraformaldehyde and subsequently hand-pealed to remove the vitelline membrane. Next, the embryos were stained with an anti-GFP antibody, allowing simultaneous imaging of GFP autofluorescence and of antibody staining in the same embryo. Figure~\ref{Figure2} shows comparison of these two probes at the surface and at the midsagittal plane of a single embryo. In both cases, the fluorescence intensity is linearly related to the protein concentration. This proportionality demonstrates that antibody stainings can be reliably used to measure relative protein concentrations in Drosophila embryos. The use of these antibodies might be decisive in quantitatively studying the subsequent gene network involved in the embryogenesis of Drosophila, particularly the gap genes and the pair-rule genes for which fluorescent fusion proteins have not yet been developed for in vivo imaging.

\section{Live imaging of \textit{Drosophila} embryos}

Genetically modified flies are kept in various cups, whose bottoms are removable, yeasted agar oviposition plates. Typically, the flies are allowed to lay for approximately one hour after changing the oviposition plate before the embryos are harvested; however, this time can be increased to ensure a larger collection of embryos on each of the plates. It is often useful to replace the oviposition plates of various cups in staggered time intervals, each lasting ~15 minutes, which will control the maximal developmental progress of each plate of embryos. This will be beneficial during the actual imaging session, as it helps to vary the time at which the embryos enter nuclear cycle 14.  
Harvested embryos are treated with pure bleach (8\% hypochloride solution) for 15 seconds in order to remove the outer chorion membrane. After rinsing, the embryos are sorted using a stereomicroscope to select for various characteristics, such as size or developmental stage. If the time allotted for oviposition is $>2$ hours, this sorting becomes important in ensuring that the embryos have not already matured passed the desired developmental stage. Once sorted, all embryos are identically oriented on an agar substrate: their anterior-posterior (head-tail) axis aligns with the y-scan of the light beam, and their dorsal-ventral (back-frong) axis aligns with the x-scan. The embryos are then mounted by carefully pushing a prepared glass slide that is coated with transparent glue onto the agar. Finally, embryos are immersed in either halocarbon oil or water, depending on the employed microscope objective. 
While planar localization of excitation remains one of the key benefits of in vivo two-photon microscopy, the excitation within this focal plane proves non-uniform due to the decrease in laser intensity with increased distance from the focal planeÕs center. In order to correct for this effect, a uniformly florescent slide is imaged to produce a flat-field correction for the later acquisitions.  
In a single imaging session $\sim100$ embryos are mounted on a slide and viewed using ScanImage (see insert, Figure~\ref{Figure1}). By saving the embryo positions in a cycle loop, each embryo may be imaged in quick succession at low resolution with 4ms/line such that their developmental progress can be monitored. The size and density of visible nuclei provide a clear indication of the embryoÕs current cell cycle. The completion of nuclear envelope degradation at the end of nuclear cycle 13 serves as a developmental marker that is used to ensure that each embryo is imaged at the same stage of development. Typically, images are acquired during early nuclear cycle 14, or 18 minutes after the above marker is reached.
Upon reaching the desired developmental stage, the imaging configuration is changed to higher resolution with 8 ms/line and three $512 \times512$ pixel frames are taken and Kalman averaged for each acquisition. Beginning with the anterior end, the embryo is imaged in three sections, with each successive image shifted 200 microns along the anterior-posterior axis of the embryo. These three images are stitched together during the data analysis to recover an image of the entire embryo in this zoomed configuration.
The laser power at the sample is adjusted between $\sim~5-40$ mW (10 mW here corresponds to $5\times10^{10}$ W/m$^2$ for a point-spread-function width of $0.5~\mu$m) with the Pockels cell. In order to assess the amount of photobleaching that occurs at a given laser power, 10 high resolution acquisitions of the previously described specifications are made in quick succession before the imaging session, and the average nuclear Bcd-GFP intensities of the images are graphed as a time series to quantify the photobleaching effect. The laser power for a given imaging session is eventually chosen such that this effect is minimized.

\section{Calculating absolute Bicoid concentration}

\begin{figure} 
   \includegraphics[width=3.2in]{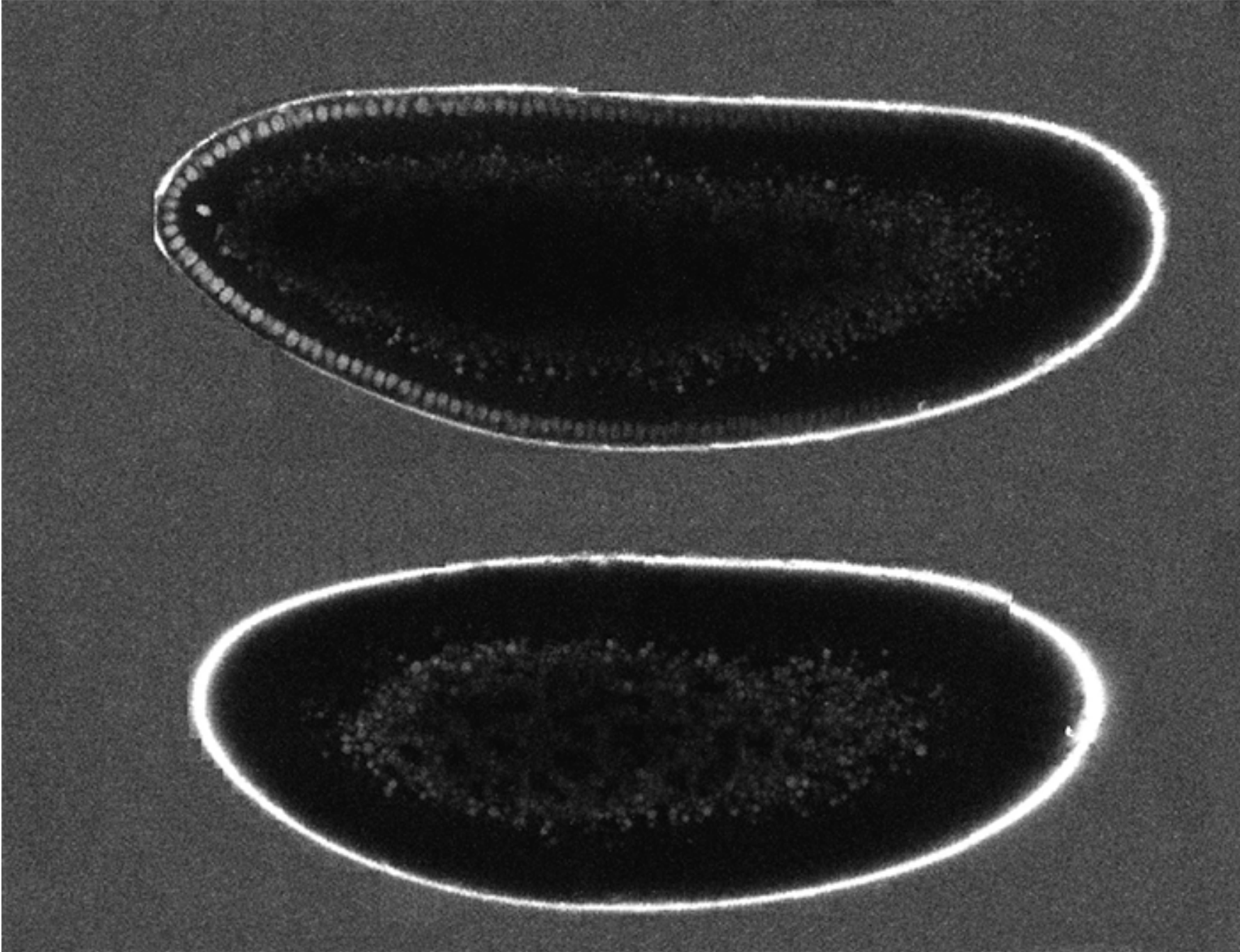}
   \centering
   \caption{Absolute concentration measurements. An embryo expressing a Bcd-GFP fusion protein (top) and a wild type embryo (bottom) immersed in a solution of 34 +/- 3nM GFP. Both embryos were taken during the same imaging session; each embryo was imaged in three pieces, which are reassembled in software. The two resulting embryo images were joined for display. (Note that embryo size difference is part of naturally varying egg sizes in wild type population.)}
   \label{Figure3}
\end{figure}

Both wild type and Bcd-GFP expressing embryos are immersed in a water solution that contains a known quantity of purified GFP molecules and imaged, as shown in Figure~\ref{Figure3}. An automated custom algorithm in Matlab (MATLAB, MathWorks, Natick, MA) identifies each nucleus within the embryo and calculates its average fluorescence intensity as follows: A mask of the embryo is obtained through a threshold that is determined by eye inspection. The original image is filtered so that each pixel within the mask is replaced by the mean intensity of a nucleus-sized disk of the corresponding pixels in the original image. From this averaged image a ring of pixels (centered on the nuclei and roughly 2 nuclear diameters wide) is created by eroding the mask. The average intensity of each pixel-segment perpendicular to the ring is determined to account for the fact that not all nuclei are located the same distance from the edge of the embryo. The algorithm, using the average values, finds the local maxima around the ring, with a minimum spacing determined by eye, to provide a rough idea of where the nuclei are. The center of the nucleus is determined as the location of the point of maximum intensity in a square of nuclear size in the averaged image centered on each peak from the ring. The intensity value for each nucleus is the average intensity of a nuclei-sized disk in the original image centered on this point. The algorithm identifies the anterior-posterior axis of the embryo and outputs a plot of the calculated intensity of each nucleus versus its position in fractional egg length along the anterior-posterior axis. The ratio of the average intensity of a nucleus-sized region in the GFP solution to the concentration of the GFP solution is used to convert the intensity values from arbitrary units into nuclear concentrations of Bcd-GFP in the embryo. To correct for background, the average intensity values of nucleus-sized regions within the wild type embryos are calculated throughout the embryo and compared across three embryos (green points with error bars in Figure 4a). This value is then subtracted from the calculated nuclear intensities to determine the absolute concentration of Bcd-GFP molecules in individual nuclei \cite{Gregor07a}.

\section{Measuring reproducibility accross embryos}

Imaging multiple live embryos using the methods described here makes it possible to measure the reproducibility of the Bcd gradient across these embryos. Initially, nuclear Bcd gradients are extracted from each embryo using the same algorithm that was used to determine absolute Bcd concentrations. These gradients are plotted together on a single graph (red dots in Figure 4a). Error bars on this graph are determined by partitioning the anterior-posterior axis into 50 bins and calculating the mean and standard deviation of the intensities in each bin (black points and error bars). The mean fluorescence background of a given image is calculated by determining the average of all intensities of the entire dataset from 90\% to 95\% egg length and subtracted from the data. Intensities are converted to absolute concentrations as described above. This allows us to quantify the reproducibility $r$, defined as $r = \sigma/\mu_{\mathrm{corr}}$ where $\sigma$ is the standard deviation and $\mu_{\mathrm{corr}}$ is the mean intensity with the background subtracted, at each location on the AP axis. In this case, the mean and standard deviation are taken over intensities from all embryos located within the given bin. The reproducibility is plotted along the AP axis with error bars determined by bootstrapping (Figure 4b) \cite{Gregor07a}. To measure cytoplasmic Bcd concentrations, a custom algorithm takes nucleus-sized disks centered at the algorithmically determined nuclei and extends each of their edges a set number of pixels (determined by eye inspection) normal to the embryo mask to create a large region around each nucleus. The cytoplasmic intensity (blue points in Figure 4a) is determined by the average intensity in this region, excluding the nucleus and a small buffer around the nucleus.

\begin{figure} 
   \includegraphics[width=3.2in]{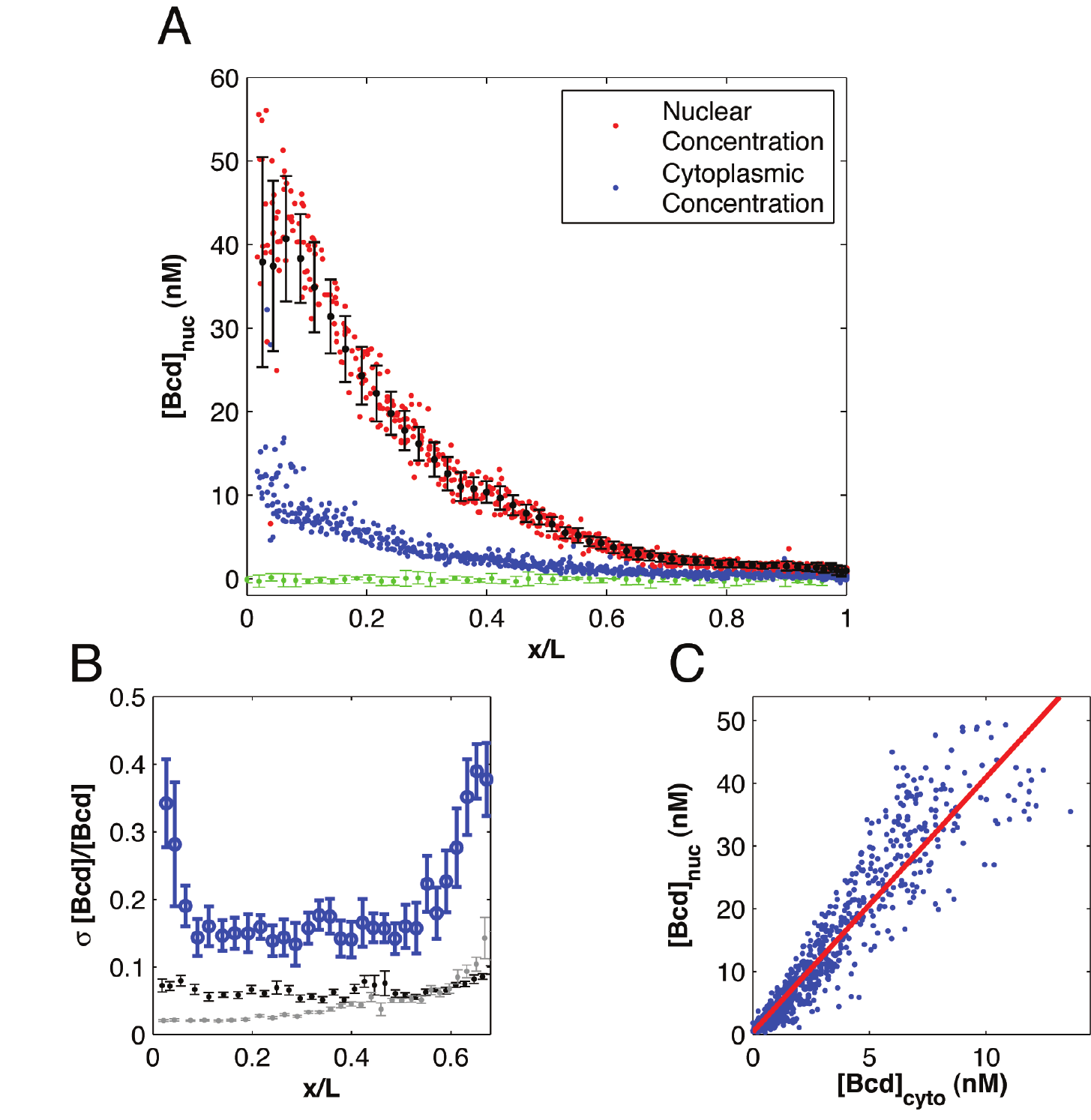}
   \centering
   \caption{Nuclear and cytoplasmic Bcd gradient measurements. \textbf{A} Nuclear and cytoplasmic Bcd-GFP profiles of 12 embryos. Each red dot represents the average concentration in a single nucleus at the midsagittal plane of the embryo (on average 80 nuclei per embryo). Each blue dot represents the average concentration in a region outside each nucleus as described in the text. All nuclei from all embryos are binned in 50 bins over which the mean and standard deviation were computed (black points with error bars). Green curve corresponds to wild type (no GFP) intensity levels. Scale at left shows raw Bcd-GFP concentration in nM. \textbf{B} For each bin in A, standard deviation divided by the mean as a function of relative egg length (blue). Error bars are computed by bootstrapping over 7 embryos. Gray and black lines show estimated contributions to measurement noise from imaging noise and focal plane adjustment noise, respectively (described in text). \textbf{C} Scatter plot for data in A shows a linear relationship between nuclear and cytoplasmic Bcd-GFP concentrations. Each dot represents a single nucleus; red line is a linear fit to all points ($R^2 = 0.89$) with a slope of $4.12\pm0.2$ (mean $\pm$ std over 5 independent datasets).}
   \label{Figure4}
\end{figure}

\section{Quantification of errors}

Four main sources of measurement noise have been determined: (1) imaging noise due to the microscope, (2) nuclear identification noise due to incorrectly centering the averaging region on the center of each nucleus, (3) focal plane adjustment noise from slight differences between the imaged plane and the actual center plane of the embryo, and (4) rotational asymmetry around the anterior-posterior axis. Imaging noise is quantified by taking five consecutive images of a small section of an embryo and by calculating the reproducibility with the mean and standard deviation taken over intensities from the five images of the same embryo. Error is introduced by photobleaching of GFP due to repeated excitation. Such photobleaching effects can be controlled by imaging an embryo repeatedly and by analyzing the significance of nuclear intensity decay with each successive image acquisition. All data presented here was obtained at a laser power where the photobleaching effect was negligible ($\sim1-2$\% during the imaging process). Nuclear identification noise is obtained by artificially displacing the algorithmically found nuclear centers. For 9 such centers forming a $3\times3$ pixel area around the algorithmically determined nuclear center a new nuclear intensity is calculated using the same averaging disk. For each nucleus a reproducibility is computed with the mean and standard deviation taken the nine locations. Focal plane adjustment noise is calculated by taking nine images, each 0.3 $\mu$m apart, with the chosen focal plane as the center image and calculating the reproducibility with the mean and standard deviation taken over intensities from the nine images of different focal planes. Error due to rotational asymmetry is estimated by comparing dorsal and ventral gradients in individual embryos to determine an upper bound on the error. The gray and black lines in Figure 4b represent the estimated contributions to noise from imaging noise and focal plane adjustment noise, respectively.

\section{Outlook}

The methods described here allow for imaging and quantification of maternal transcription factors in living fly embryos. Over the next decade, this approach is likely to be extended to zygotic transcription factors and other concentration measurements of the proteome, hopefully leading to a complete dynamic description of the early fly patterning cascade. Further, it should be fairly straightforward to extend this approach to related species, provided that transgenic lines can be generated to incorporate a fusion protein containing a GFP derivative. Particularly insightful would be to visualize multiple GFP derivatives of different colors in the same embryo, all tagging different transcription factors of the same small regulatory network. Cross-correlation analyses of data generated from such embryos would give us direct access to the underlying structure of the network and effectively help elucidate its design principles.

\begin{acknowledgments}
We are grateful to our collaborators for aiding in the collection and analysis of some of the data presented here, and for comments on the chapter: William Bialek, Shawn Little, Feng Liu, Alistair McGregor, David Tank, Stephan Thiberge and Eric Wieschaus. Our work has been supported by Princeton University, NIH grant R01 GM077599 and HHMI.
\end{acknowledgments}

\end{document}